\documentclass[epj,twocolumn]{webofc}
\usepackage[varg]{txfonts}  
 \woctitle{ISVHECRI 2018}
\usepackage{amsmath}
\usepackage{amsfonts}
\usepackage{graphicx}
\usepackage{mathrsfs}
\usepackage{comment}
\usepackage{verbatim}   
\usepackage{color}      
\usepackage{subfigure}  
\usepackage{hyperref}   
\newcommand{\Slash}[1]{{\ooalign{\hfil#1\hfil\crcr\raise.167ex\hbox{/}}}}

\newcommand{\beq}{\begin{equation}}  \newcommand{\eeq}{\end{equation}}
\newcommand{\bef}{\begin{figure}}  \newcommand{\eef}{\end{figure}}
\newcommand{\bec}{\begin{center}}  \newcommand{\eec}{\end{center}}
\newcommand{\non}{\nonumber}  \newcommand{\eqn}[1]{\begin{equation} {#1}\end{equation}}
\newcommand{\laq}[1]{\label{eq:#1}}  

\newcommand{\Eq}[1]{Eq.(\ref{eq:#1})}

\newcommand{\eq}[1]{(\ref{eq:#1})}

\newcommand{\ab}[1]{\left|{#1}\right|}
\newcommand{\vev}[1]{ \left\langle {#1} \right\rangle }

\def\o{\over}
\def\a{\alpha}
\def\b{\beta}

\def\d{\delta}
\def\e{\epsilon}
\def\f{\phi}

\def\h{\theta}

\def\l{\lambda}
\def\m{\mu}
\def\n{\nu}
\def\p{\psi}

\def\s{\sigma}
\def\t{\tau}

\def\G{\Gamma}

\def\L{\Lambda}

\def\ol{\overline}
\def\tl{\tilde}
\def\*{\dagger}

\def\({\left(}
\def\){\right)}

\def\diag{\mathop{\rm diag}\nolimits}

\def\O{\mathcal{O}}

\newcommand{\AND}{~{\rm and}~}
\newcommand{\EV}{ {\rm ~eV} }

\newcommand{\MEV}{ {\rm ~MeV} }
\newcommand{\GEV}{ {\rm ~GeV} }

\begin{document}

\title{Highly-boosted dark matter and cutoff for cosmic-ray neutrino  through neutrino portal}
\author{ Wen Yin\thanks{\email{yinwen@kaist.ac.kr}}}

\institute{ Institute of High Energy Physics, Chinese Academy of Sciences, Beijing 100049,  China \and Department of Physics, Korea Advanced Institute of Science and Technology, 
291 Daehak-ro, Yuseong-gu, Daejeon 34141, Republic of Korea~\thanks{After the talk, the affiliation was changed from IHEP to KAIST.}
}

\abstract{ We study the cutoff for the cosmic-ray neutrino, set by the scattering with cosmic background neutrinos into dark sector particles through a 
neutrino portal interaction. We find that a large interaction rate is still viable, when the dark sector particles are mainly coupled to the $\tau-$neutrino, so that the 
neutrino mean free path can be reduced to be $\O(10)~$Mpc over a wide energy range.
If stable enough, the dark sector particle, into whom most of the cosmic-ray neutrino energy is transferred, can travel across the Universe and reach the earth. The dark sector particle can carry the energy as large as $\O({\rm EeV})$ if originates from a cosmogenic neutrino.}

\maketitle
\section{Introduction}

The cosmogenic neutrinos are known as a ``guaranteed" flux of cosmic ray neutrinos, 
which are generated through the photo-pion production, where the interaction between ultra-high cosmic rays (UHECRs) and ambient photon backgrounds~\cite{Beresinsky:1969qj},
This interaction was first proposed to sets a cutoff for UHECRs and the cutoff as well as the UHECRs have been already observed~\cite{Greisen:1966jv,
Zatsepin:1966jv}.
However the IceCube observatory, whose one of the purposes is to detect the cosmogenic neutrinos, has not yet observed neutrinos with energy $\gg 10^{6}\GEV$. 
The Glashow resonance around $6$~PeV is also not observed \cite{Glashow:1960zz}.

These non-observations may relate to the neutrality of the neutrino.  
In particular, the neutrinos can interact with a dark sector, 
whose participants can carry some hidden charge, so that the lightest particle is stable to explain the dark matter i.e. there could be neutrino portal interaction relevant to dark matter~\cite{Kaplan:2009ag, Bertoni:2014mva}. 
It was pointed out that a cutoff for the cosmic-ray neutrino can be set through its scattering with cosmic background neutrinos (C$\nu$B)  into the dark sector particles through the neutrino-portal interaction by using an effective theory approach~\cite{Yin:2017wxm}. (For other models or other scattering processes affecting cosmic-ray neutrinos, see e.g. Refs \cite{Ioka:2014kca, Ng:2014pca, Ibe:2014pja,Cherry:2014xra}.)
Thus, the Universe could be opaque to the cosmic-ray neutrino.

Here we study the propagation of the cosmic-ray neutrinos with Dirac neutrino portal interaction, where the right-handed neutrinos are Dirac-type particles. This is a renormalizable model.
The reason we consider Dirac-type right-handed neutrinos rather than the Majorana-type is to suppress the lepton number violating effect.
We show that there are viable parameter regions, especially for the $\tau$-neutrino ($\n_\tau$) portal interaction, 
that the cosmic-ray neutrinos get scattered with the C$\nu$B into the 
particles in the dark sector before they 
reach the earth. As a result, the neutrino flux is transferred into the dark sector particles which relaxes the tension from the non-observations. 
In particular, there are viable parameter regions, that the neutrino flux is highly-suppressed, which might be difficult to be detected in the observatories.
Instead, a large fraction of the viable region may be tested in collider experiments. 
Since the energy of the cosmic-ray neutrinos is transferred into the dark matter, there could be flux of highly-boosted dark matter in the universe. They reach 
the earth instead of the cosmic-ray neutrinos which may be tested in various neutrino observatories.

\section{Dirac neutrino portal interaction}

Let us consider the following interaction with lepton number symmetry as one example.
\beq
\laq{lag}
\d{\cal L}_{\rm int}=-\sum_{i=e,\mu,\t}\( {y_{Ri}}{{H}\ol{\n}_{Ri} \hat{P}_L L_i }  + \tl{y}_i  \ol{\n}_{Ri} \hat{P}_R \p\f_i\)+{\rm h.c.}
\eeq
Here, $H$ is the standard model (SM) Higgs boson and $L_i$ is the left handed leptons in flavor basis where $i=e,\m,\t$.
$\p, \n_{Ri} \AND \f_i$ are right-handed (Dirac) neutrino, a Majorana fermion, and leptonic scalars, respectively. 
These Yukawa matrices are diagonalized without loss of generality.

The mass terms of these fields are given by 
\beq
\laq{mass}
\d{\cal L}_{\rm mass}=-\sum_{i=e,\mu,\t}{\(m_{\f ij}^2 \f^\*_{i} \f_j  +M_{Rij} \ol{\n}_{Ri} {\n}_{Rj} +{1\o 2} M_\p \ol{ \p}\p\)}+{\rm h.c.},
\eeq
where we have defined the lepton number conserving mass squares $m_{\f_{ij}}^2$, Dirac masses for the right-handed neutrinos $M_{Rij}$, and the Majorana mass term $M_\p$ for fermion $\p$.
The symmetry allows the potentials for $\f_i$ of the form,
\beq
\laq{pot}
V= {1\o 4 }\sum_{i,j,k,l} \l_{i j k l} \f_i^\*  \f_j^\* \f_k \f_l+   \sum_{i,j} \f_i^\* \f_j (\e_{ij} \ab{H}^2-\e_{ij}  v^2).
\eeq
Here, $v\simeq 174\GEV$ is the vacuum expectation value of the Higgs field.
The first term lets a leptonic scalar decay into the lightest scalars in the dark sector, if kinematically allowed.
We will assume $\e_{ij}$ of the second term is negligible to define the neutrino portal interaction. 
Notice that this is rarely generated radiatively when $\tl{y}_i  y_j$ are small enough.

At the lepton number symmetric limit, the SM neutrinos are massless. 
With some small soft breaking terms for lepton number, such as $\tl{M}_{Rij}\ol{\n^c_{Ri}} {\n}_{Rj}$, and $B_{ij} \f_i \f_j$, the neutrino masses are generated, through inverse seesaw mechanism \cite{Mohapatra:1986bd}, or radiatively \cite{Yin:2017wxm}.
In both cases, a $Z_2$ symmetry remains from the lepton number breaking. 
The $Z_2$ symmetry makes the lightest one of $\f_i, \p$ stable, and allows it to be dark matter.
The small lepton number breaking term does not change the following discussion, and we will work on with the symmetric limit.

For simplicity, hereafter we will take 
\beq
m_{\f ij}^2= \diag{\{m_{\f_e}^2,m_{\f_\mu}^2,m_{\f_\tau}^2\}}, M_{Rij}= \diag{\{M_{R_e},M_{R_\mu},M_{R_\tau}\}}.
\eeq 
 As discussed later, the experimental constraints on the $\n_{R e}$ and $\n_{R \m}$ are severer than the ones for $\n_{R\t}$ in the interesting parameter ranges. 
 Thus, we will assume $y_{R\mu} \AND y_{Re} $ are negligible and focus on the interaction for $\tau$ sector for simplicity.

\section{Cutoff for neutrino flux}

Now, consider the impact of the portal interaction to the propagation of neutrino in the Universe.

The neutrino $\a$ (in mass eigen state) mixes with the $\n_{R\t}$, with mixing angle $\simeq U_{\t}\times (U_{\rm PNMS})_{\t \a}$. Thus the cosmic-ray neutrino scatters with the C$\nu$B through the mixing,
 \eqn{\laq{nnan} \n_{\a}+ \n_\b^{\rm C\n B}/\ol{\n}_\b^{\rm C\n B}  \rightarrow \p+\p , \f_\t+\f_\t /\f^*_\t.} 
 The total scattering cross section with center of mass energy $E_{\rm cm}$ is given by
\beq
\laq{GZK}
\s_{\a\b}  v_{\rm rel} \simeq   \theta\left({
1-{2m_{\f_\t}\o E_{\rm cm}}}\right) R_{\a\b} \frac{
1
}{32 \pi  {E_{\rm cm}^2}} \log \left(\frac{{E_{\rm cm}}+\sqrt{{E_{\rm cm}^2}-4{m_{\f_\t}^2}}}{{E_{\rm cm}-\sqrt{{E_{\rm cm}^2}-4{m_{\f_\t}^2}}}}\) \eeq
where we have defined 
\beq
R_{\a\b}=\ab{(U_{\rm PNMS})_{\a \t}}^2 \ab{(U_{\rm PNMS})_{\b \t}}^2 \ab{U_\t}^4 \ab{\tl{y}_\t}^4 
\eeq
and assumed $M_\p=m_{\f_\t}$ for illustrative purpose;  $v_{\rm rel} $ is the relative velocity between the two neutrinos at the center of mass frame; $U_i\equiv {y_i v\o M_{Ri}}$ is the mixing angle between the left and right handed neutrinos; We have taken the average for the target neutrino and anti-neutrino for the crosssection;
The unit stepfunction $\h$ represents the momentum conservation.
The energy of the cosmic ray $E_\nu$ at our frame is related with the center of mass energy as \beq E_{\rm cm}\simeq \sqrt{2\(\sqrt{\ab{\bf p}^2+m_{\n \a}}E_{\rm \n}-\ab{{\bf p}}E_{\rm \n} \cos\h \)}\eeq with neutrino mass $m_{\n \a}$ and momentum $\bf p$ of a C$\nu$B neutrino.

\begin{figure}[!t]
\sidecaption
\includegraphics[width=7cm,clip]{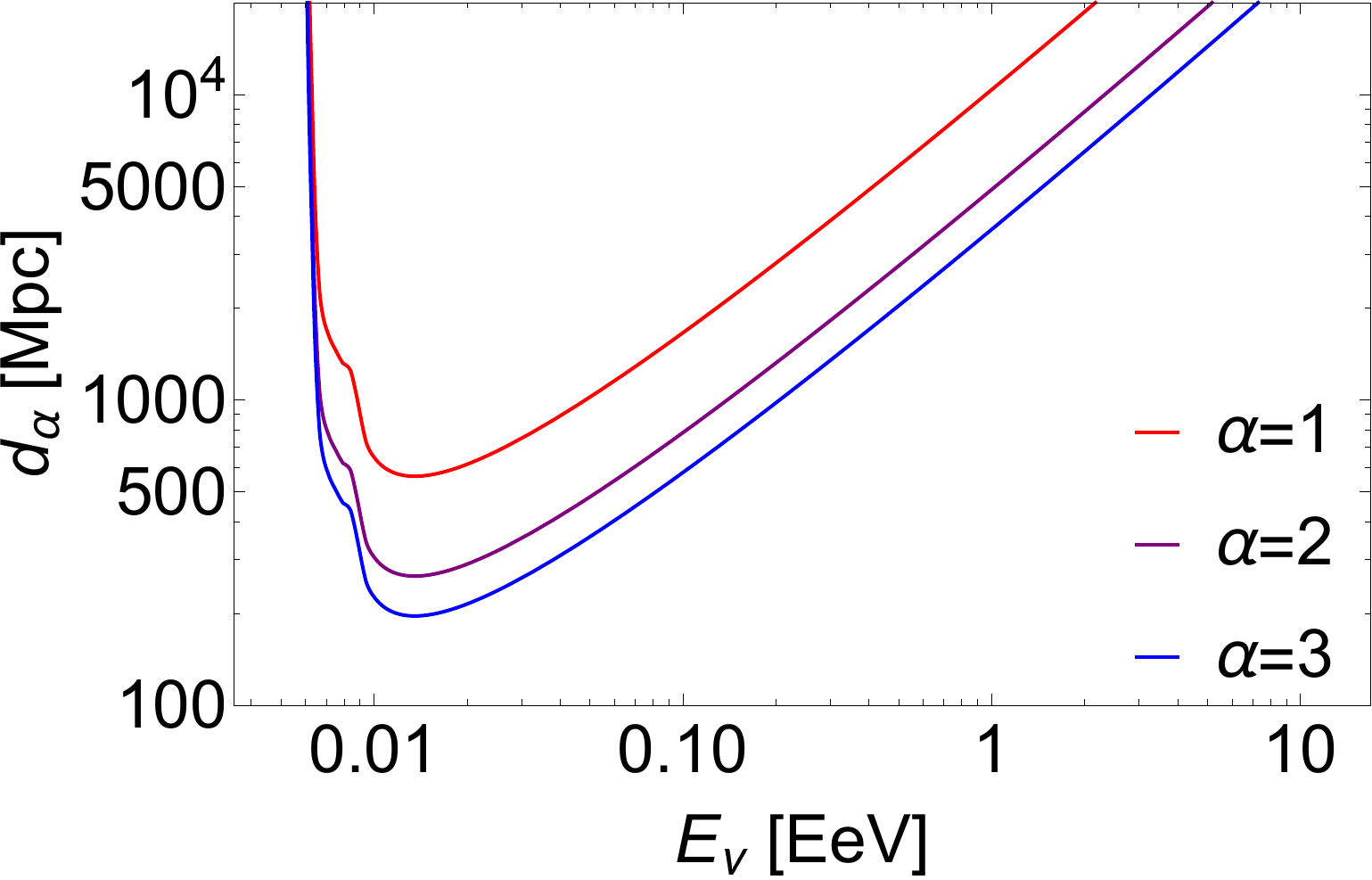}
\caption{ The mean free path for the neutrino in mass eigenstate $\a$ as a function of energy. 
We take $M_\p=m_{\f_\t}=15\MEV$ and  $\ab{U_\t \tl{y}_\t}=0.4$. 
}
\label{fig:1}
\end{figure}

The interaction rate of the neutrino $\n_\a$ is obtained by taking thermal average of the previous cross section, 
\beq
\laq{MFP}
\G_{\rm \a}(E_\n, T_\n ) = \sum_\b \vev{\s_{\a\b} v_{\rm rel} n_{\n \b}} .
\eeq
Here, $n_{\n \a}({\bf p})=2/(e^{\ab{\bf p}/T_\n}+1)$ is the neutrino distribution in $\rm C \n B$. $\vev{}$ denotes the thermal average.
The mean free path of neutrino is defined by \beq d_\a (E_\n , T_\n )\equiv {1 \o \G_{\a}(E_\n , T_\n )}.\eeq
The numerical result for $d_\a(E_\n ,T_\n^{\rm now})$ is presented Fig. \ref{fig:1},  where $T_\n^{\rm now}\simeq 1.69\times10^{-4} \EV$ is the current temperature of the $\rm C\n B$.
Here and hereafter, we take the normal mass ordering case with lightest neutrino mass $0.05\EV$ and the Dirac phase $\d_{13}=-\pi/2$.  
The numerical result can be approximately read as
\beq
\G_{\a} \sim \( 1 ~{\rm  Gpc}\)^{-1} \ab{\tl{y}_\t U_\t \o 0.3}^4 \( {300 \MEV \o E_{\rm cm}}\)^2 \h(E_{\rm cm}-m_{\f_\t}/2).
\eeq
The stepfunction implies that the mean free path of neutrino with 
\beq 
E_\nu \gtrsim {m_{\phi_\t}^2\o \max{\{ m_{\n{\rm lightest}}, T_\nu}\}}\eeq
becomes smaller than $\O(\rm Gpc)$, when
 \beq
|\tl{y}_\tau U_\t| \gtrsim \O(0.1).
\eeq
By giving the neutirno masses of $\O(0.001-0.1)\EV$, this implies that the neutrino with $E_\n =10^{6-10}\GEV$ is possible to be scattered before it travels over $\O(1)$~Gpc for $m_{\phi_\t} \simeq \O(10-100)\MEV.$\footnote{  The lower bound of $m_{\phi \t}$ is from the cosmological constraints on dark radiation.} This can give explanation of non-observation of cosmogenic neutrino as well as the Glashow resonance.\\

 \begin{figure}[!t]
\sidecaption
\includegraphics[width=7cm,clip]{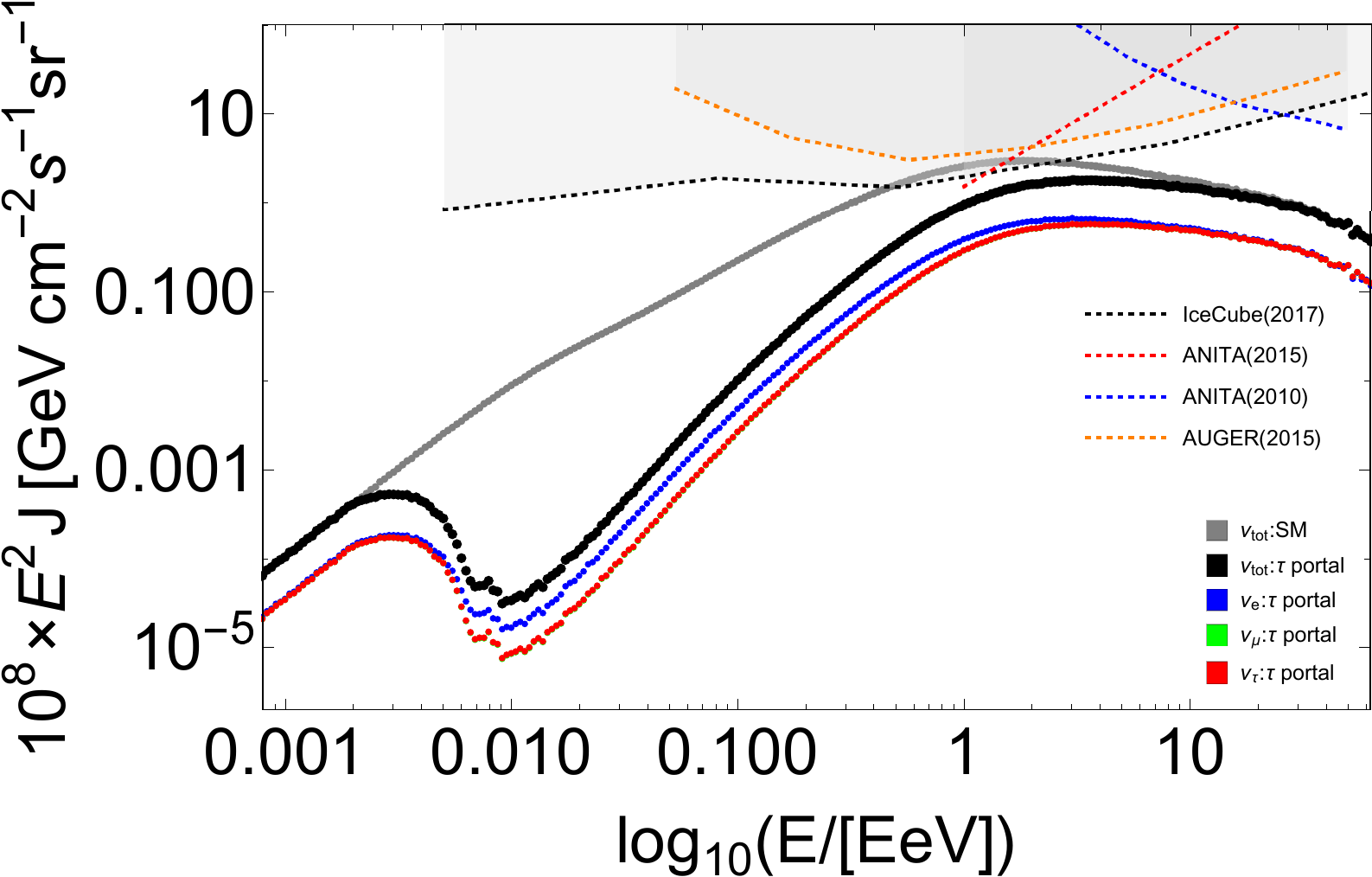}
\caption{ The flux $\times E^2$ for the total neutrinos (black), $\nu_e$ (blue) ,$\nu_\mu$ (Green), $\nu_\tau$ (red) as a function of energy. For comparison, flux $\times E^2$ of total neutrinos within the SM is also shown (gray). We take $M_\p=m_{\f_\t}=15\MEV$ and $\ab{U_\t \tl{y}_\t}=0.4$. The constraints (shaded regions) at 90\% CL are adapted from \cite{Gorham:2010kv, Schoorlemmer:2015afa, Aab:2015kma,
Aartsen:2017mau}.
}
\label{fig:GZK}
\end{figure}

 \begin{figure}[!t]
\sidecaption
\includegraphics[width=7cm,clip]{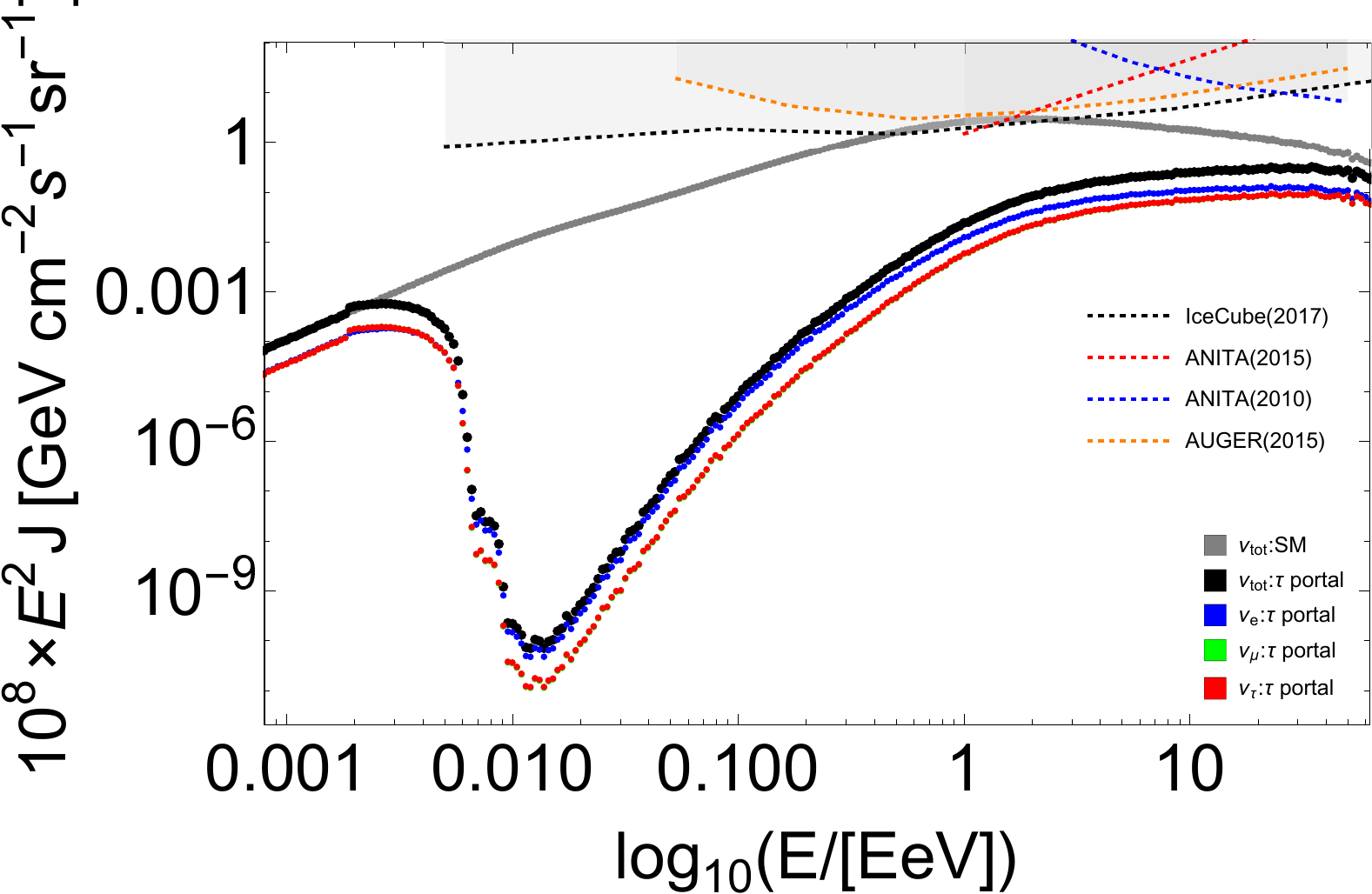}
\caption{ Same as Fig.\ref{fig:GZK}, but $\ab{U_\t \tl{y}_\t}=1$. 
}
\label{fig:GZK2}
\end{figure}
Now let us perform a numerical estimation on the neutrino flux at the earth. 
We assume an original cosmogenic neutrino flux and that the produced $\f_\t,\f_\t^*\AND \p$ does not lead to energetic secondary neutrinos for a moment.
This might be the case that $\f_\t,\f_\t^*\AND \p$ soon decay into the other dark sector fields.
A case with secondary neutrinos will be studied in the next section.

The number density of the cosmogenic neutrinos are produced at a rate $n_{\n_e}: n_{\n_\mu}: n_{\n_\t} \simeq 1: 2: 0 $ of flavor through photo-pion interaction.
The number density in the mass basis is given by $n_{\n_\a}= \sum_i (U^*_{\rm PNMS})_{\a i}(U_{\rm PNMS})_{\a i} n_{\n_i}$ and $n_{\n_1}: n_{\n_2}: n_{\n_3} \sim 1:1:1$. Since the free propagating length $d_\a \gtrsim \O({\rm Mpc})$ is much longer than the neutrino oscillation scale, the interference between different neutrinos in mass basis disappears due to the rapid oscillation. As a result, the neutrinos travel in the mass basis with the ratio $\sim 1:1:1$ kept.

When the traveling distance becomes $\gtrsim d_\a$ the neutrino flux is gradually transferred into the  dark sector through scattering between the neutrino and $\rm C\nu B$.
A neutrino $\a$ emitted at red shift $z=z_s$ travels until now at a survivability of,
\beq
\laq{sub}
R_\a (E_{\rm now}, z_s) =e^{- \int_{0}^{z_s}{dz \ab{{dt \o dz}}  {\G_{\a}(E_{\rm now}(1+z), T_\n (1+z)) }}},
\eeq 
where $E_{\rm now}$ is the energy measured in the current universe, and we have taken account of the effect for the redshift for the energy and $T_\nu$. Here $\ab{dt\o dz}\equiv \( {\( 1+z\)} \sqrt{\Omega_M (1+z)^3 +\Omega_\L}\)^{-1}$. Thus, the neutrino in a flavor basis $i$ reaches to the earth at a probability,
\beq
\tl{R}_i=R_\a(E_{\rm now}, z_s) \ab{(U_{\rm PNMS})_{i\a}}^2.
\eeq

Using {\tt CRPropa 3}\cite{Batista:2016yrx} we have made a numerical simulation on the cosmogenic neutrino source, assuming the observed UHECR are purely protons. 
For each cosmogenic neutrino, we calculate the survivability \eq{sub} from its profile given by {\tt CRPropa 3}.  As a result, we obtain the neutrino flux as in Figs. \ref{fig:GZK} and \ref{fig:GZK2} for the effective coupling $\ab{\tl{y}_\t U\t}=0.4$ and $1$, respectively.
In the numerical simulation, we use one of the best fit parameter given in Ref. \cite{Ahlers:2010fw}:
the proton source of energy $E_p$ is set to have a power law distribution $\propto (E_p)^{-2.49}$ between $10^{17.5}\EV<E_p<10^{21}\EV$ for $z<z_{\rm max}=2$, the cosmic evolution rate is represented by $(1+z)^{3.5}$ until $z=z_{\rm max}$. 
Also shown is the current bound from the experiments of IceCube, AUGER, and ANITA. We found that the cosmogenic neutrino flux to the earth can be significantly reduced due to the neutrino portal interaction. 

\begin{figure}[!t]
\sidecaption
\includegraphics[width=7cm,clip]{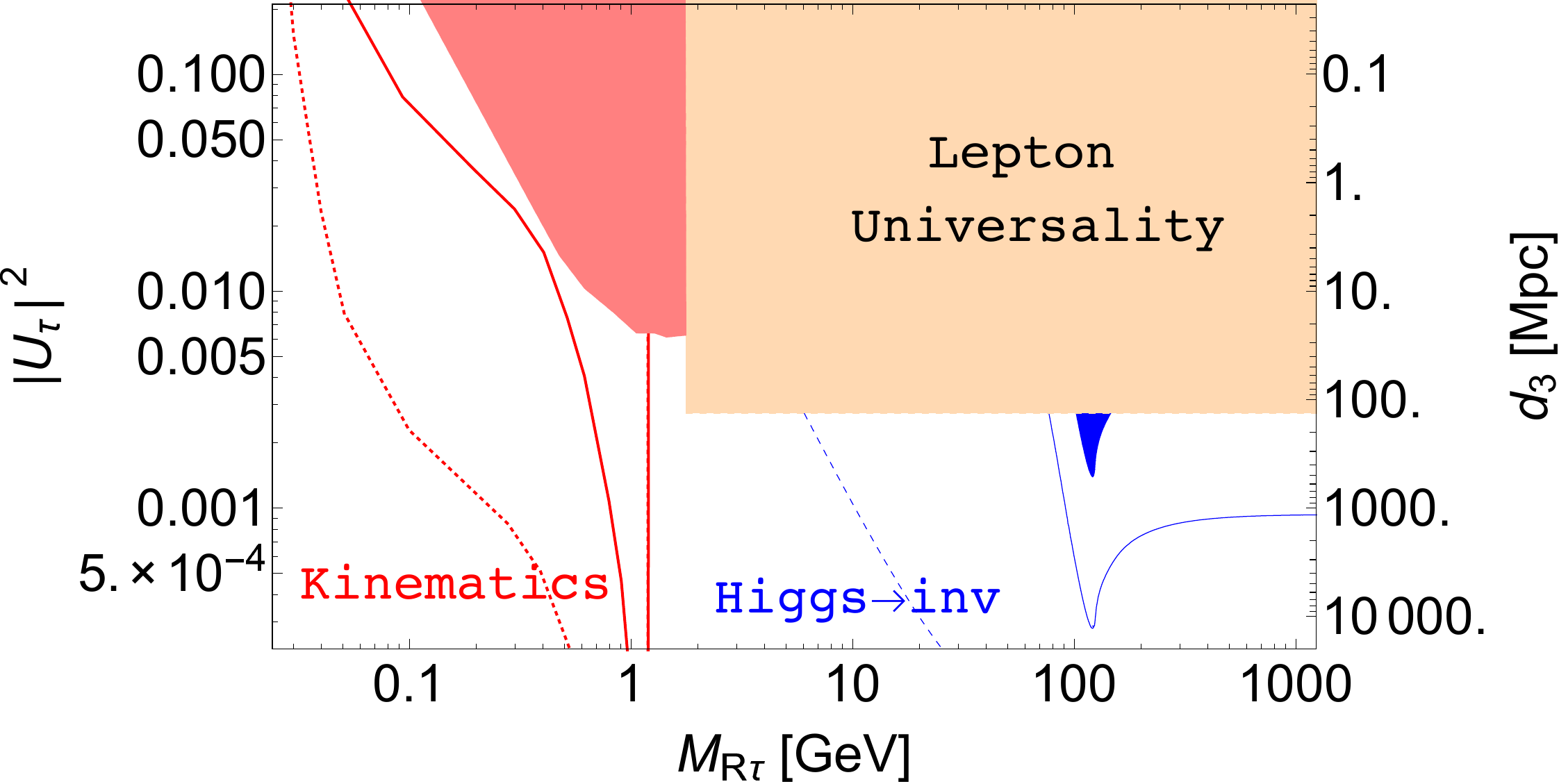}
\caption{ The viable region for $\tau$-neutrino portal interaction. The mean free path of heaviest neutrino with $E_\n=1$\,EeV is given in the right-handed side.  We have taken $M_{R\t}>m_{\f_\t}\simeq M_\p=15\MEV$ and $\tl{y}_\t=4\pi$. The colored region might be excluded.  }
\label{fig:2}
\end{figure}

\begin{figure}[!t]
\sidecaption
\includegraphics[width=7cm,clip]{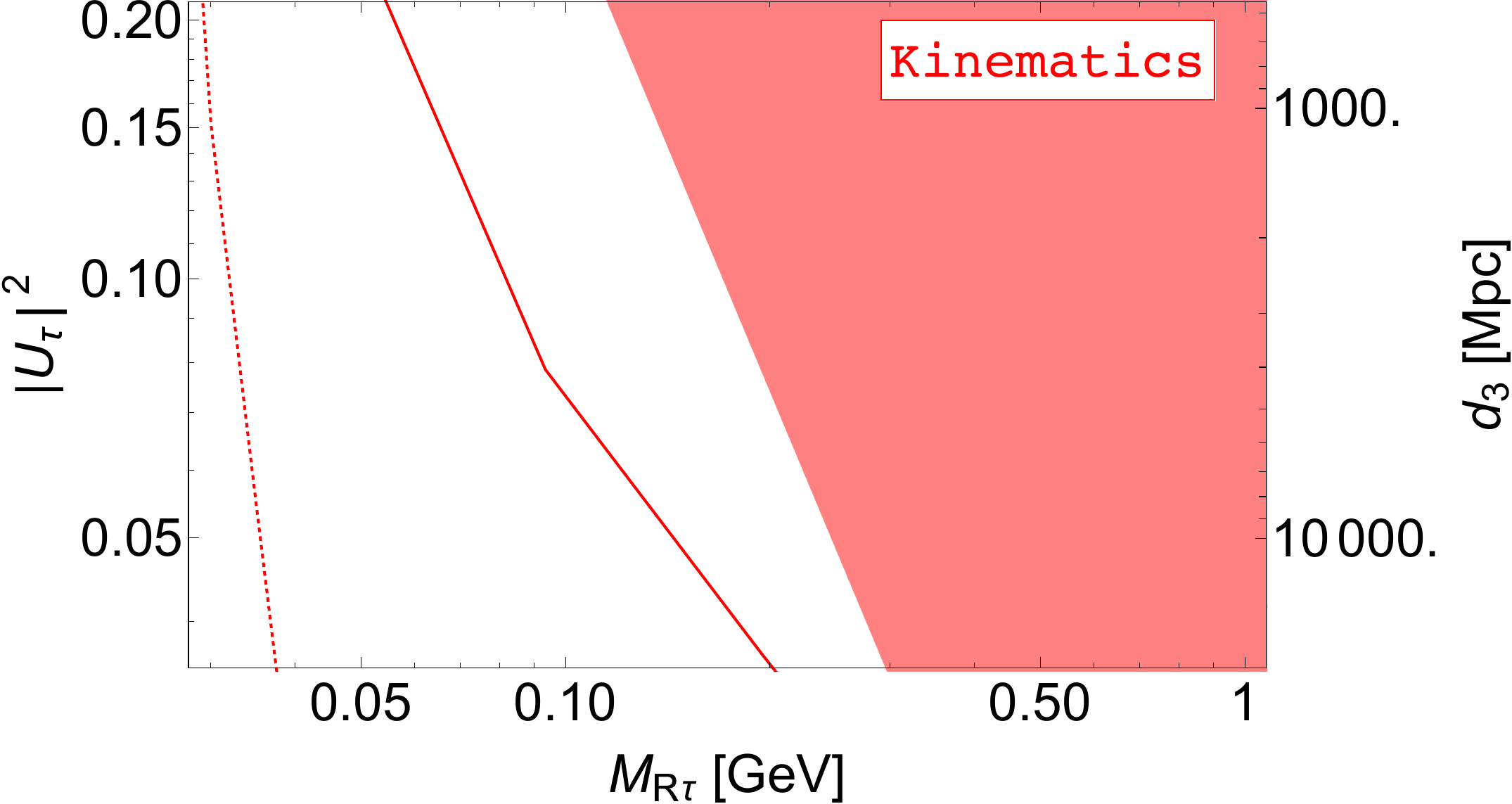}
\caption{ Same as Fig. \ref{fig:2} except for $\tl{y}_\t=1$.}
\label{fig:3}
\end{figure}

In fact, there are allowed region for the $\n_\t$ portal interaction with $\ab{\tl{y}_\t U_\t} \lesssim \O(1)$. 
Furthermore, a large portion of the allowed region can be tested in future from collider experiments. 
In Figs.~\ref{fig:2} and \ref{fig:3} we show the viable/testable parameter region for this model, and the mean free path at the right end.
When $\n_{R\tau}$ is heavier than $\tau$, it only decays into lighter mass eigenstate with weak current ratio ${g_\t \o g_{\m,e}} =\sqrt{1-\ab{U_\tau}^2}$.  
The constraints on the ratio are given by ${g_\t \o g_{\m}} = 1.0001 \pm 0.0014$ and ${g_\t \o g_{e}} =1.0029\pm 0.0015$ \cite{Lusiani:2016ofc}, and we have combined them
to find the lepton universality bound (orange band) \beq \ab{U_\t}\lesssim 0.0027~ (99\%{\rm CL}) ~~(M_{R\t}>m_\t).
\eeq
When $M_{R\tau}$ is much smaller than $m_\t$, the neutrino in the decay product is represented by the flavor eigenstate and the lepton universality bound does not apply.
When $M_{R\t}$ is slightly lighter than the $\tau$ lepton, the kinematics of the visible decay products of $\t$ are different, which could be distinguished. This indirectly constrains the parameter region (Red band: $95\%{\rm CL}$ limit) \cite{Helo:2011yg,
Kobach:2014hea}.  In particular, it was discussed in Ref. \cite{Kobach:2014hea}, the future B-factories could test the scenario with kinematic measurement of semi-leptonic $\t$ decay (red solid line: conservative,
red dashed line: optimistic).

The Higgs boson can decay into neutrinos and the dark sector fields or the right-handed neutrino. Such decay enhances the branching ratio of the Higgs boson to missing energy. 
The contribution to the decay rate of our scenario can be calculated as,
\begin{align}
\G_{H\rightarrow {\rm missing}}& \simeq   {\ab{y_{R\t }}^2\o 16\pi} m_h \(1-\frac{\ab{M_{R\t}}^2}{m_h^2}\)^2 \theta{(1-\ab{M_{R\t}}/m_h)}\non\\
&+\frac{\ab{\tl{y}_{\t}^2 y_{ R\t}^2}}{512 \pi ^3 {m_h}^3} \left(2 M_{R\t}^2 (\ab{M_{R\t}}^2-{m^2_h}) \log \left(\left| \frac{{m_h}^2}{M_{R\t}^2}-1\right| \right)\right. \non \\& \left.  -{m_h}^4+2 {m_h}^2 M_{R\t}^2\right).
\end{align}
The tree contribution for the process $H\rightarrow \n_\t+\ol{\n}_{R\t}$ is given in the first raw. 
In particular, we have included the process $H \rightarrow \n_\t + \f_\t+\p$, which becomes important when $M_{R\t}>m_h$. We have made an analytical continuation in the calculation of the three-body decay so that the leading loop correction for the two-body decay is also included. 
We have neglected the mass of $\f_{\t}$ and $\p$, whose interesting range for us is much smaller than the Higgs boson mass.
When $4 \lesssim \tl{y}_{\t} \lesssim 4\pi$, which may suggest $\f_\t$ is like a pion, 
the region affect the cosmic-ray neutrino propagation may also be tested from the Higgs boson decay in the LHC which is proposed to measure the branching ratio at a precision of $0.05$ (blue solid line) \cite{Peskin:2013xra}.\footnote{Blue shaded region may be excluded \cite{Aad:2015pla, Khachatryan:2016whc}.}
The future lepton colliders may reach $\sim 0.001$ (blue-dashed line) \cite{CEPC, 
CEPC2,Asner:2013psa,dEnterria:2016sca,Abramowicz:2016zbo}.\footnote{We note that $\ab{U_\tau} < 0.42$ from the $\tau-\mu$ neutrino oscillation with matter effects~\cite{Abe:2014gda}.}

Notice that we have taken the mass range that $\n_{R\t}$ decays into $\f_\t$ and $\p$. 
In the case the decay channel is forbidden, $\n_{R\t}$ could have a much longer lifetime and decay into the SM particles. 
This leads to severer constraints from beam dump experiments, as well as cosmology (see Refs. \cite{Astier:2001ck,Orloff:2002de} ). 
However, there is still allowed region at $M_{R\t}\simeq \O(10)\MEV$ to affect the cosmic-ray neutrino flux.  

When the neutrinos portal is through $\n_\m$ or $\n_e$ the experiment constraints becomes severer \cite{Bertoni:2014mva}. The measurement on meson decays sets stringent constraints and almost excludes all the viable region supressing the cosmic-ray neutrino with $M_{Ri}\lesssim m_K\sim 500 \MEV$. 
Above the kaon mass, there are constraints from lepton universality as the $\t$ case. 
When $M_{Ri}\gtrsim m_K$ with $\tl{y}_{i}>\O(1)$, there are still viable regions.
The prediction for the Higgs boson decay, which we have calculated, holds.

\section{Highly boosted dark matter}
Since $\phi_i$ and $\p$ are the only $Z_2$ odd particles, the lightest one among them is stable and 
could be the dark matter. 
This gives a distinguishable and interesting prediction of our scenario: the highly-boosted dark matter whose energy originates from the cosmogenic neutrinos.  
There are two cases for this scenario: (a) the dark matter is directly produced through the interaction 
\eq{nnan}, (b) the dark matter is produced through the cascade decay of $\f_\t \AND \p$. 
The discussion in the previous section can correspond to the case (b),
 where the decay products are assumed to rarely interact with $\rm C \n B$ and to be without introducing secondary energetic neutrinos.

\begin{figure}[!t]
\sidecaption
\includegraphics[width=7cm,clip]{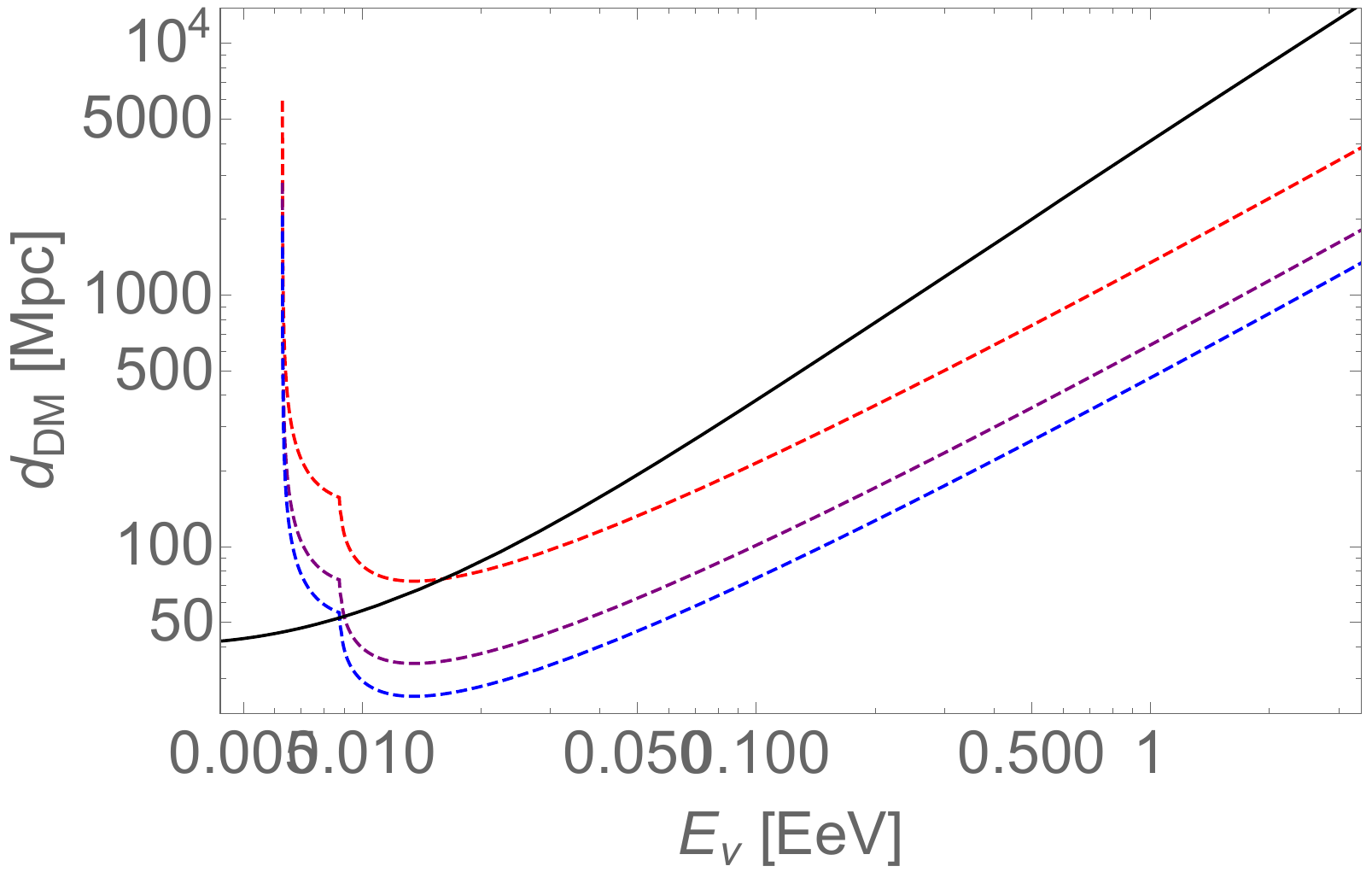}
\caption{ The energy loss length for dark matter (black solid). $m_{\f_\t}\simeq M_\p=15\MEV$ and $\ab{\tl{y}_\t U_\t}=0.5$ are taken. For comparison the mean free path of neutrinos are also shown (see Fig. \ref{fig:1}).}
\label{fig:4}
\end{figure}

Now let us consider case (a). 
For simplicity, suppose that $0< M_{\p}-m_{{\phi_\t}}\ll m_{\phi_\t} $, and $\f_e \AND \f_\m$ are much heavier.  
In this case, the produced dark matter, $\f_\t$, has energy $\sim E_\n/2$ (or produced $\p$ decays to $\f_\tau$ with energy $E_\n/2$ and a soft $\tau$ neutrino.). 
The dark matter travels across the universe for a long distance. It loses energy dominantly through the scattering with the C$\nu$B,\footnote{We have assumed that the right-handed neutrino is heavy enough that the process to $\f_\t +\nu_{\t_R}$ is forbidden.}
\beq 
\laq{DMscat}
\f_\t + \nu_\t^{\rm C\nu B}/\ol{\nu}_\t^{\rm C\nu B} \rightarrow \f_\t +\nu_\a/\ol{\nu}_\a.
\eeq
Since this is an elastic scattering process, the energy-loss rate becomes more important than the interaction rate.
The energy loss rate for the interaction is given by
\begin{align}
\laq{Eloss}
&\G_{{\rm DM},\a} \equiv {d\o d t} \log{E}\\ & \simeq \ab{(U_{\rm PNMS})_{\t\a}}^2 \sum_{\b=1,2,3}\vev{\s_{\f_\t \b} v_{\rm rel} n_{\n \b} \(1-{E^{\rm f}_{\rm DM}\o E^{\rm i}_{\rm DM} }\)} \ab{(U_{{\rm PNMS}})_{\tau \b}}^2.
\end{align}
 where $E_{\rm DM}^{\rm i}$ and $E_{\rm DM}^{\rm f}$ are the energy of the dark matter in the initial and final states; $\s_{\s_{\f_\t \a}}$ is the scattering crosssection of \eq{DMscat}. The energy loss length is given by
 \beq
 d_{\rm DM} \equiv {1 \o  \sum_{\a} \G_{\rm DM,\a}}.
 \eeq
 This is shown in the Fig. \ref{fig:4}.
 $d_{\rm DM}$ is slightly longer than $d_{\a}$ at $E_{\rm cm}\gg m_{\f_\t},M_\p$. 
This is due to the helicity suppression. Analytically one finds
\beq
\G_{\rm DM,\a}  \sim {1\o 3} \(1+\log{\( E^2_{\rm cm} \o \ab{M_{\p}}^2\)}\)^{-1}   \G_{\a}.
\eeq
which implies a suppression of factor several for $\G_{\rm DM,\a}$ compared with $\G_{\a}$. 
As a result, the dark matter at high energy produced through \Eq{DMscat} is easier to reach the earth than a cosmogenic neutrino.

\begin{figure}[!t]\sidecaption
\includegraphics[width=7cm,clip]{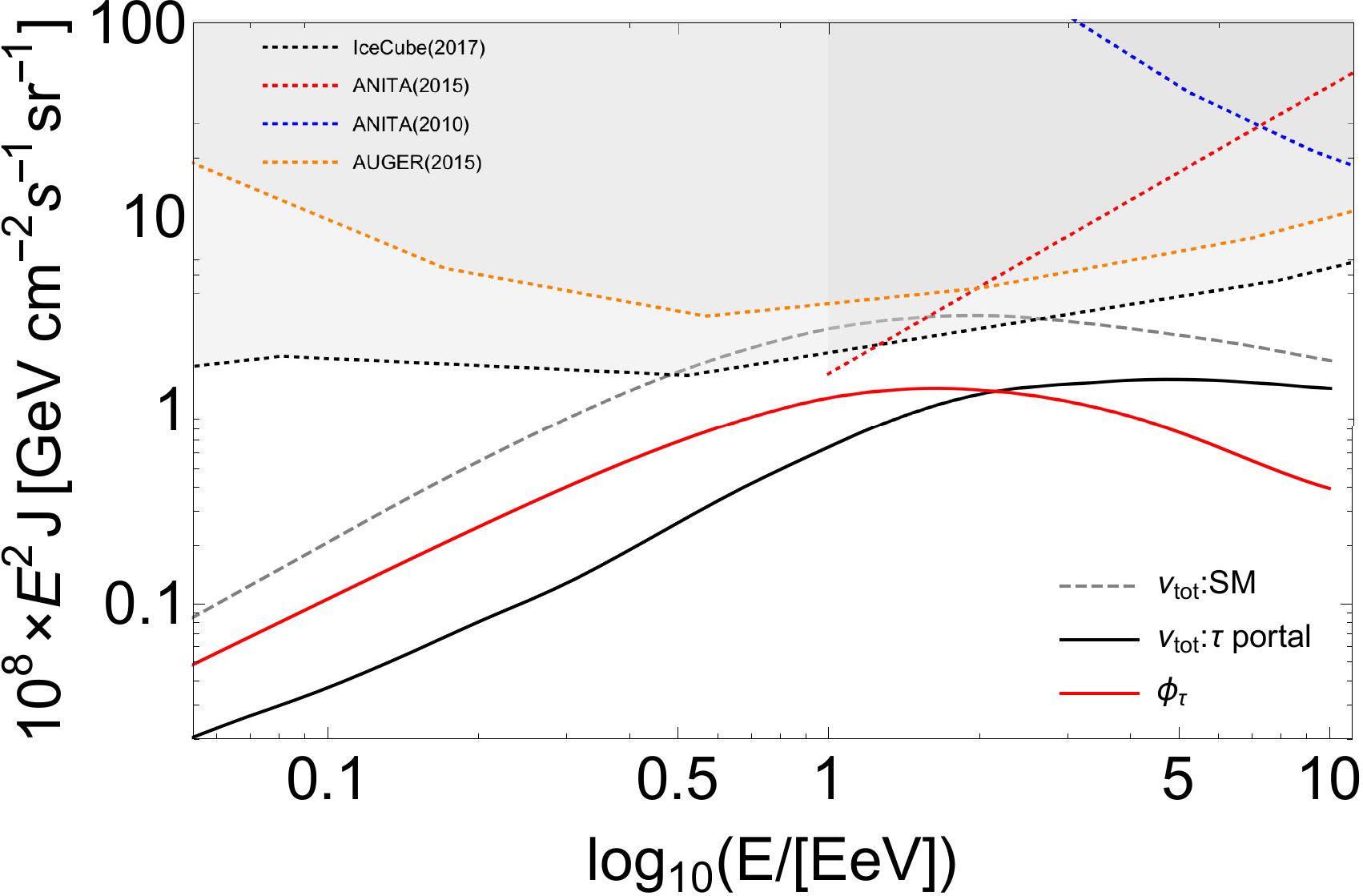}
\caption{ The predicted fluxes of neutrinos (black solid line) and dark matter (red solid line).  $m_{\f_\t}\simeq M_\p=15\MEV$ and $\ab{\tl{y}_\t U_\t}=0.5$ are taken. }
\label{fig:5}
\end{figure}

We perform a numerical simulation on the evolution of the number density of neutrinos and produced 
dark matter. 
The differential equation is given by,
\begin{align}
\non &{\partial \o \partial t}  J_{\n\a} (E_\n) \simeq H(z)  {\partial \o \partial { E_\n }} {\(E_\nu J_{\n\a}\)} -\G_{\a}  J_{\n \a} + J_{\rm source}\\&~~~~~~~~~~~~~~~~+\int{d{E_{\f_\t}} {{d \s_{\rm \f_\t \a} \o d{E_{\n}} } v_{\rm rel} n_{\rm C\n B }}J_{\f_\t}}(E_{\f_\t}), 
\laq{res1} \\
\non &{\partial \o \partial t}J_{\f_\tau}(E_{\f_\tau})\simeq H(z) {\partial \o \partial { E_{\f_\t}}} (E_{\f_\t} J_{\f_\t})- \G_{\f_\t} J_{\f_\t}\\&~~~~~~~~~~~~~~~~ +\int{d{E_{\n}}  \sum_{\a}{{{{d \s_{\rm \a} \o d {E_{\f_\t}} }v_{\rm rel} n_{\rm C\n B}}J_{\n\a }}}(E_\n) } \non \\ 
&~~~~~~~~~~~~~~~~+\int{d{\tl{E}_{\f_\t}} \sum_{\a}{{{d \s_{\rm \f_\t \a } \o d {E_{\f_\t}} } v_{\rm rel} n_{\rm C\n B}}J_{\f_\tau }}(\tl{E}_{\f_\t})}, \laq{res2}\end{align} 
Here, $J_{...}\equiv {\partial {n_{...}} \o \partial \log{E}} $; $H(z)$ is the Hubble parameter at $z$; $\G_{\rm \f_\t}(E_\n, T_\n ) = \sum_\b \vev{\s_{\f_\tau \b} v_{\rm rel} n_{\n \b}}$; $J_{\rm source}$ represents the cosmogenic neutrino number density emitted at red shift $z(t)$, which is fitted from {\tt CRPropa 3}; for simplicity we have assumed that the neutrino and anti-neutrinos are in same distribution. The terms with integrals represent the re-scattering process. In the equation, $E$ and $T_\n$ are related to the current value by $E=E_{\rm now} (1+z(t))$ and $T_\n=(1+z(t)) T_\n^{\rm now}$, at the time $t$. 
The numerical result is given in Fig. \ref{fig:5} for $\ab{\tl{y}_\t U_\t} = 0.5$. One finds that the peak flux of dark matter can be as large as the one for the original cosmogenic 
neutrino for $E=\O(10^9) \GEV$.

\paragraph{Discussion}

The highly boosted dark matter scatters with a nucleon as $\f_\t+N\rightarrow \p +\t/\n_\t + N$. The cross section is of order, 
$
\laq{DMneuc}
\s_{\f_\t N} \sim {1\o 16\pi^2} \ab{\tl{y}U_\t}^2 \times \s_{\n_\t N}
$
where $\s_{\n \t N}$ is the $\n_\t$-$N$ scattering cross section of $\n_\t +N\rightarrow \t + N$ in SM 
and $1\o 16\pi^2$ represents the phase space suppression. 
Since this is $\O(10^{-3})-\O(10^{-1})$ suppressed to $\s_{\n_\t N}$, 
Since the earth is much more transparent for the dark matter than a SM neutrino,
the events for the highly boosted dark matter can be distinguished from the ordinary cosmogenic neutrinos from how long it travels within the earth.  It may be tested or might be already detected in the ANITA experiment (c.f. Refs.~\cite{Gorham:2018ydl, Cherry:2018rxj}).

The coupling of $\ab{\tl{y}_\t U_\t}=\O(0.1)$ suggests a too large annihilation cross section for $\f_\t$ to 
get a correct thermal relic abundance. One needs another dominant dark matter candidate.\footnote{This may be the inflaton (e.g. Refs.~\cite{Daido:2017wwb, Daido:2017tbr}) or some superpartners~(e.g. Refs.\cite{Shimizu:2015ara,
Yin:2016shg,
NGH,
Yin:2016pkz,
Yanagida:2018eho}).}
$\p$, if is extended into a Dirac fermion, can be asymmetric dark matter \cite{Kaplan:2009ag}. Interestingly, the parameter region for addressing small-scale structure issues also has $y_\tau U_\tau=\O(0.1)$ and $m_{\rm DM}=\O(10-100)\MEV$ \cite{Bertoni:2014mva}, which may be coincident with the parameter region suppressing cosmic-ray neutrinos and highly boosted dark matter. 
The clarification of the coincidence will be our future study. 

\section{Conclusions}

We have shown that a cutoff for the cosmic-ray neutrino can be set by the scattering with cosmic background neutrinos through a neutrino portal interaction. 
In particular, a large interaction rate is still allowed for $\tau-$neutrino portal interaction which is being/to be tested in the on-going/future collider experiments. 
Highly-boosted dark matter can reach the earth instead of the cosmic-ray neutrinos and may be tested in the near future and might be already detected in the ANITA experiment.


\begin{thebibliography}{99}

  
\bibitem{Beresinsky:1969qj} 
  V.~S.~Berezinsky and G.~T.~Zatsepin,
  Phys.\ Lett.\  {\bf 28B}, 423 (1969).
  
  
  
\bibitem{Greisen:1966jv} 
  K.~Greisen,
  Phys.\ Rev.\ Lett.\  {\bf 16}, 748 (1966).


  
\bibitem{Zatsepin:1966jv} 
  G.~T.~Zatsepin and V.~A.~Kuzmin,
  JETP Lett.\  {\bf 4}, 78 (1966)
  [Pisma Zh.\ Eksp.\ Teor.\ Fiz.\  {\bf 4}, 114 (1966)].
  
  
  
  
\bibitem{Glashow:1960zz} 
  S.~L.~Glashow,
  Phys.\ Rev.\  {\bf 118}, 316 (1960).





\bibitem{Kaplan:2009ag} 
  D.~E.~Kaplan, M.~A.~Luty and K.~M.~Zurek,
  Phys.\ Rev.\ D {\bf 79}, 115016 (2009)
  [arXiv:0901.4117 [hep-ph]].
  
\bibitem{Bertoni:2014mva} 
  B.~Bertoni, S.~Ipek, D.~McKeen and A.~E.~Nelson,
  JHEP {\bf 1504}, 170 (2015)
  [arXiv:1412.3113 [hep-ph]].




\bibitem{Yin:2017wxm} 
  W.~Yin,
  arXiv:1706.07028 [hep-ph].


\bibitem{Ioka:2014kca} 
  K.~Ioka and K.~Murase,
  PTEP {\bf 2014}, no. 6, 061E01 (2014)
  [arXiv:1404.2279 [astro-ph.HE]].
  

  
   \bibitem{Cherry:2014xra} 
  J.~F.~Cherry, A.~Friedland and I.~M.~Shoemaker,
  arXiv:1411.1071 [hep-ph].



  
  
\bibitem{Ng:2014pca}
  K.~C.~Y.~Ng and J.~F.~Beacom,
  Phys.\ Rev.\ D {\bf 90} (2014) no.6,  065035
   Erratum: [Phys.\ Rev.\ D {\bf 90} (2014) no.8,  089904]
[arXiv:1404.2288 [astro-ph.HE]].
  
 \bibitem{Ibe:2014pja} 
  M.~Ibe and K.~Kaneta,
  Phys.\ Rev.\ D {\bf 90}, no. 5, 053011 (2014)
  [arXiv:1407.2848 [hep-ph]].



\bibitem{Batista:2016yrx} 
  R.~Alves Batista {\it et al.},
  JCAP {\bf 1605}, no. 05, 038 (2016)
  [arXiv:1603.07142 [astro-ph.IM]].
  

\bibitem{Schoorlemmer:2015afa} 
  H.~Schoorlemmer {\it et al.},
  Astropart.\ Phys.\  {\bf 77}, 32 (2016)
  [arXiv:1506.05396 [astro-ph.HE]].


\bibitem{Aartsen:2017mau} 
  M.~G.~Aartsen {\it et al.} [IceCube Collaboration],
  arXiv:1710.01191 [astro-ph.HE].



\bibitem{Aab:2015kma} 
  A.~Aab {\it et al.} [Pierre Auger Collaboration],
  Phys.\ Rev.\ D {\bf 91}, no. 9, 092008 (2015)
  [arXiv:1504.05397 [astro-ph.HE]].

  
\bibitem{Gorham:2010kv} 
  P.~W.~Gorham {\it et al.} [ANITA Collaboration],
  Phys.\ Rev.\ D {\bf 82}, 022004 (2010)
  Erratum: [Phys.\ Rev.\ D {\bf 85}, 049901 (2012)]
  [arXiv:1003.2961 [astro-ph.HE], arXiv:1011.5004 [astro-ph.HE]].
  
  \bibitem{Ahlers:2010fw} 
  M.~Ahlers, L.~A.~Anchordoqui, M.~C.~Gonzalez-Garcia, F.~Halzen and S.~Sarkar,
  Astropart.\ Phys.\  {\bf 34}, 106 (2010)
  [arXiv:1005.2620 [astro-ph.HE]].
  

\bibitem{Mohapatra:1986bd} 
  R.~N.~Mohapatra and J.~W.~F.~Valle,
  Phys.\ Rev.\ D {\bf 34}, 1642 (1986).
\bibitem{Malinsky:2005bi} 
  M.~Malinsky, J.~C.~Romao and J.~W.~F.~Valle,
  Phys.\ Rev.\ Lett.\  {\bf 95}, 161801 (2005)
  [hep-ph/0506296].


\bibitem{Hooper:2004jc} 
  D.~Hooper, A.~Taylor and S.~Sarkar,
  Astropart.\ Phys.\  {\bf 23}, 11 (2005)
  [astro-ph/0407618].



\bibitem{Lusiani:2016ofc} 
  A.~Lusiani [BaBar Collaboration],
  EPJ Web Conf.\  {\bf 118}, 01018 (2016).
 
 
\bibitem{Helo:2011yg} 
  J.~C.~Helo, S.~Kovalenko and I.~Schmidt,
  Phys.\ Rev.\ D {\bf 84}, 053008 (2011)
  [arXiv:1105.3019 [hep-ph]].
 
\bibitem{Kobach:2014hea} 
  A.~Kobach and S.~Dobbs,
  Phys.\ Rev.\ D {\bf 91}, no. 5, 053006 (2015)
  [arXiv:1412.4785 [hep-ph]].
 

\bibitem{Peskin:2013xra} 
  M.~E.~Peskin,
  arXiv:1312.4974 [hep-ph].

\bibitem{Aad:2015pla} 
  G.~Aad {\it et al.} [ATLAS Collaboration],
  JHEP {\bf 1511}, 206 (2015)
  [arXiv:1509.00672 [hep-ex]].
  

\bibitem{Khachatryan:2016whc} 
  V.~Khachatryan {\it et al.} [CMS Collaboration],
  JHEP {\bf 1702}, 135 (2017)
  [arXiv:1610.09218 [hep-ex]].
  


\bibitem{CEPC} 
CEPC-SPPC Study Group,\\http://cepc.ihep.ac.cn/preCDR/main\_ preCDR.pdf
\bibitem{CEPC2} 
CEPC-SPPC Study Group,\\
http://cepc.ihep.ac.cn/preCDR/Pre-CDR\_final\_20150317.pdf

\bibitem{Asner:2013psa} 
  D.~M.~Asner {\it et al.},
  arXiv:1310.0763 [hep-ph].

\bibitem{dEnterria:2016sca} 
  D.~d'Enterria,
  arXiv:1602.05043 [hep-ex].


\bibitem{Abramowicz:2016zbo} 
  H.~Abramowicz {\it et al.},
  arXiv:1608.07538 [hep-ex].
  


\bibitem{Abe:2014gda} 
  K.~Abe {\it et al.} [Super-Kamiokande Collaboration],
  Phys.\ Rev.\ D {\bf 91}, 052019 (2015)
  [arXiv:1410.2008 [hep-ex]].

  
  
\bibitem{Orloff:2002de} 
  J.~Orloff, A.~N.~Rozanov and C.~Santoni,
  Phys.\ Lett.\ B {\bf 550}, 8 (2002)
  [hep-ph/0208075].

\bibitem{Astier:2001ck} 
  P.~Astier {\it et al.} [NOMAD Collaboration],
  Phys.\ Lett.\ B {\bf 506}, 27 (2001)
  [hep-ex/0101041].

  
\bibitem{Gorham:2018ydl} 
  P.~W.~Gorham {\it et al.} [ANITA Collaboration],
  arXiv:1803.05088 [astro-ph.HE].


\bibitem{Cherry:2018rxj} 
  J.~F.~Cherry and I.~M.~Shoemaker,
  arXiv:1802.01611 [hep-ph].


\bibitem{Daido:2017wwb} 
  R.~Daido, F.~Takahashi and W.~Yin,
``The ALP miracle: unified inflaton and dark matter,''
  JCAP {\bf 1705}, no. 05, 044 (2017)
  [arXiv:1702.03284 [hep-ph]].
  
\bibitem{Daido:2017tbr} 
  R.~Daido, F.~Takahashi and W.~Yin,
``The ALP miracle revisited,''
  JHEP {\bf 1802}, 104 (2018)
  [arXiv:1710.11107 [hep-ph]].
  
  \bibitem{Shimizu:2015ara} 
  Y.~Shimizu and W.~Yin,
  Phys.\ Lett.\ B {\bf 754}, 118 (2016)
  [arXiv:1509.04933 [hep-ph]].
%

\bibitem{Yin:2016shg}
  W.~Yin and N.~Yokozaki,
  Phys.\ Lett.\ B {\bf 762}, 72 (2016)
  [arXiv:1607.05705 [hep-ph]].
  
\bibitem{NGH}
  T.~T.~Yanagida, W.~Yin and N.~Yokozaki,
  JHEP {\bf 1609}, 086 (2016)
  [arXiv:1608.06618 [hep-ph]].
%


\bibitem{Yin:2016pkz} 
  W.~Yin,
  Chin.\ Phys.\ C {\bf 42}, no. 1, 013104 (2018)
  [arXiv:1609.03527 [hep-ph]].
  
\bibitem{Yanagida:2018eho} 
  T.~T.~Yanagida, W.~Yin and N.~Yokozaki,
  JHEP {\bf 1804}, 012 (2018)
  [arXiv:1801.05785 [hep-ph]].

  \end{thebibliography}
\end{document}